\documentclass[9pt]{extarticle}
\usepackage[utf8]{inputenc}
\usepackage{spconf}
\usepackage{hyperref}

\usepackage{graphicx}
\usepackage{musixtex}
\usepackage{subcaption}
\usepackage{paralist}
\usepackage{pgfplots}
\usepackage{tikz}
\usepackage{verbatim}
\usepackage{listings}

\pgfplotsset{compat=1.12}
\usepgfplotslibrary{statistics}

\title{BandNet: A Neural Network-based, Multi-Instrument Beatles-Style MIDI Music Composition Machine}

\name{
Yichao Zhou$^{,1,2}\qquad$
Wei Chu$^{,1}\qquad$
Sam Young$^{1,3}\qquad$
Xin Chen$^{1}\qquad$
}
\address{
$^1$ Snap Inc. 63 Market St, Venice, CA 90291, \\
$^2$ Department of EECS, University of California, Berkeley,\\
$^3$ Herb Alpert School of Music, University of California, Los Angeles \\
zyc@berkeley.edu, wei.chu@snap.com, samyoungmusic@gmail.com, xin.chen@snap.com 
}

\begin{document}

\maketitle

\begin{abstract}
In this paper, we propose a recurrent neural network (RNN)-based MIDI music composition machine that is able to learn musical knowledge from existing Beatles' songs and generate music in the style of the Beatles with little human intervention. In the learning stage, a sequence of stylistically uniform, multiple-channel music samples was modeled by a RNN. In the composition stage, a short clip of randomly-generated music was used as a seed for the RNN to start music score prediction. To form structured music, segments of generated music from different seeds were concatenated together. To improve the quality and structure of the generated music, we integrated music theory knowledge into the model, such as controlling the spacing of gaps in the vocal melody, normalizing the timing of chord changes, and requiring notes to be related to the song's key (C major, for example). This integration improved the quality of the generated music as verified by a professional composer. We also conducted a subjective listening test that showed our generated music was close to original music by the Beatles in terms of style similarity, professional quality, and interestingness. Generated music samples are at 
\href{https://goo.gl/uaLXoB}{https://goo.gl/uaLXoB}.

\end{abstract}

\section{Introduction}


Automatic music composition has been an active research area for the last several decades, and researchers have proposed various methods to model many different kinds of music.  \cite{quick2014kulitta,ebciouglu1988expert,whorley2013multiple,hiller1979experimental,eppe2015computational} used rules and criteria developed by professional musicians to generate songs. These methods usually relied  heavily on the input of music experts, hand-crafted rules, consistent intervention during the process of composition, and fine-tuning the generated music in the post-processing stage. Although the quality of the composed music may be quite satisfactory, the composition process can be time-consuming and the composed music can be biased toward a particular style. Recently, agnostic approaches that do not depend on expert knowledge have been emerging \cite{hadjeres2016deepbach}.  Instead of relying on music experts, these methods employ a data-driven approach to learn generalizable theory and patterns from existing pieces of music, and this approach has proven to be effective. For example, \cite{allan2005harmonising,kaliakatsos2014probabilistic} trained a hidden Markov model from music corpora and \cite{hadjeres2016style} modeled polyphonic music from the perspective of the graphic model. With the recent progress made in deep learning, there have been many research efforts that have tried to compose music using neural networks:  \cite{yang2017midinet} used a deep convolutional network to generate a melody conditioned on the chords found in each measure; \cite{makris2017combining} generated the drum pattern for songs using a RNN \cite{hochreiter1997long}; \cite{hadjeres2016deepbach,huang2016counterpoint,liang2016bachbot} described RNN approaches to modeling and harmonizing Bach-style polyphonic music; and \cite{chu2016song} proposed a multi-layer RNN to model pop music by encoding drum and chord patterns as one-hot vectors.



While most of the aforementioned machine-learning methods were able to generate music in some categories, we found that it is challenging to use them in modeling songs by the Beatles. The musical style of the Beatles is characterized by catchy vocal melodies, unique chord progressions, and an upbeat, energetic sound. The standard instrumentation of the Beatles is vocals, two electric guitars, bass, drums, and occasional piano. One difficulty of replicating the Beatles' music is that the component parts depend on each other but have different characteristics. For example, the bass line is often monophonic while the guitar chords are polyphonic, and the guitar chords are likely to contain certain notes found in the bass part. The model needs to be able to generate different instrumental parts within a uniform musical structure. In addition, the style of the musical features often changes between songs. For example, many Beatles' songs use monophonic vocal melodies while others use polyphonic, two-part vocal melodies, and the chords in the Beatles' music can be played by either piano or guitar, each of which use different chord spacings. All of these variations are challenging to model. Moreover, the Beatles are known for using complex harmonies that can be difficult to classify, with the added complication that certain chords may be incomplete or missing one or more of their component parts. Thus it may not be appropriate to encode the chord progression aspect of the music as one-hot vectors, as they treat two similar harmonies differently.

To overcome these difficulties, we introduce BandNet, a RNN-based, Beatles-style multi-instrument music composition machine. The proposed approach will be explained in Section \ref{sec:method} and compared with other approaches in Section \ref{sec:exp}. 



%




\section{Methods} \label{sec:method}

\subsection{Data Representation}
Our BandNet uses MIDI files as input and output and utilizes the same data processing pipeline from Magenta \cite{magenta14}.  For each Beatles' song, we consider the three most important channels: the vocal melody, guitar chords, and bass part.  All the channels are allowed to be polyphonic, to maximize the flexibility of the model.

In our dataset we include only songs that use a 4/4 time signature, which means that a quarter note is felt as the beat, and each measure (a.k.a one bar, a short segment of music whose boundaries are shown by vertical bar lines in the score) has four beats. It is reasonable to discretize note lengths into sixteenth notes. We call the duration of a sixteenth note a \emph{step}. Therefore, each measure is discretized into 16 steps and each beat is discretized into 4 steps.

Because a song may be played by different instruments with different pitch ranges, we first transpose the pitch by octave so that the average pitch of each channel in each song is as close as possible to the global pitch average of that channel. Next, we transpose each song by -5 to 6 semitones to augment the training data by 12 times so that it is able to generate music in all possible keys. Other approaches, such as transposing each song to the same key, C major for example, do not work well for the Beatles' music because we have yet to find a reliable way to detect the key of each song.

\subsection{Score Encoding}
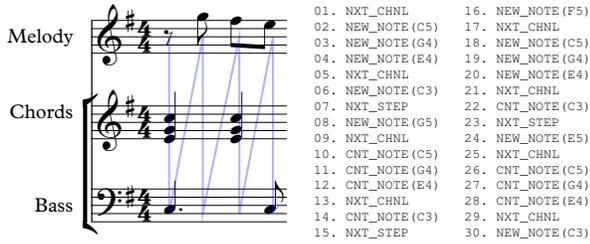
\begin{figure}
    \centering
    \begin{subfigure}[b]{0.48\linewidth}
        \centering
        \begin{tikzpicture}
          \node[anchor=south west,inner sep=0] (image) at (0,0) {\includegraphics[width=.9\linewidth]{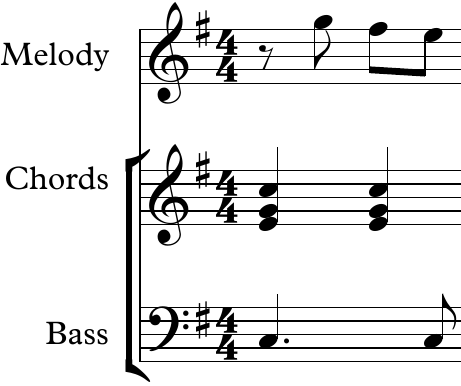}};
          \begin{scope}[x={(image.south east)},y={(image.north west)}]
              \coordinate (a1) at (0.58, 0.85);
              \coordinate (a2) at (0.58, 0.1);
              \coordinate (b1) at (0.70, 0.85);
              \coordinate (b2) at (0.70, 0.1);
              \coordinate (c1) at (0.83, 0.85);
              \coordinate (c2) at (0.83, 0.1);
              \coordinate (d1) at (0.95, 0.85);
              \coordinate (d2) at (0.95, 0.1);
              \draw[->,blue, thick,opacity=0.3] (a1) -> (a2) -> (b1) -> (b2) -> (c1) -> (c2) -> (d1) -> (d2);
          \end{scope}
      \end{tikzpicture}
      \caption{\small A sheet music example.  The scan line is marked in blue.}
    \end{subfigure}%
    ~ 
    \begin{subfigure}[b]{0.48\linewidth}\centering%
\lstset{%
  basicstyle={\ttfamily\tiny},%
  columns=fullflexible,%
  keepspaces=true,%
}%
\begin{lstlisting}
01. NXT_CHNL       16. NEW_NOTE(F5)
02. NEW_NOTE(C5)   17. NXT_CHNL
03. NEW_NOTE(G4)   18. NEW_NOTE(C5)
04. NEW_NOTE(E4)   19. NEW_NOTE(G4)
05. NXT_CHNL       20. NEW_NOTE(E4)
06. NEW_NOTE(C3)   21. NXT_CHNL
07. NXT_STEP       22. CNT_NOTE(C3)
08. NEW_NOTE(G5)   23. NXT_STEP
09. NXT_CHNL       24. NEW_NOTE(E5)
10. CNT_NOTE(C5)   25. NXT_CHNL
11. CNT_NOTE(G4)   26. CNT_NOTE(C5)
12. CNT_NOTE(E4)   27. CNT_NOTE(G4)
13. NXT_CHNL       28. CNT_NOTE(E4)
14. CNT_NOTE(C3)   29. NXT_CHNL
15. NXT_STEP       30. NEW_NOTE(C3)
\end{lstlisting}\vspace{-8pt}
\caption{\small The encoded sequence of the sheet music on the left.}
    \end{subfigure}
    \vspace{8pt}
    \caption{\small An example showing how we encode an excerpt from \emph{I Want to Hold Your Hand (1964)}.  Notes are quantized to eighth notes rather than sixteenth notes for demonstration purposes.}
\label{fig:encode}
\end{figure}

BachBot \cite{liang2016bachbot} and Magenta \cite{magenta14} convert polyphonic MIDI music into a sequence of symbols so that RNN can be used to model the probabilistic distribution of such a sequence. We expand their encoding scheme to music with multiple channels.


Figure \ref{fig:encode} gives an example showing how we encode the music score.  We create a new type of symbol \texttt{NXT\_CHNL}, along with the three existing categories: \texttt{NEW\_NOTE}, \texttt{CNT\_NOTE}, and \texttt{NXT\_STEP}.  The strategy is to scan the score in a left to right (time dimension), top to bottom (channel dimension), zig-zag fashion.  Each time we meet a note during the scan, we will first check whether it is a new note or a continuation of a previous note (e.g., the second sixteenth interval of an eighth note).   We will then either emit a \texttt{NEW\_NOTE} or a \texttt{CNT\_NOTE} symbol depending on the case, followed by the pitch of that note.  When a channel is polyphonic, the note with higher pitch will always be in front of the notes with lower pitch according to this strategy.  When the scan line comes across the boundary of a channel, we will emit a \texttt{NXT\_CHNL} symbol, and when the scan line comes across a time step, we will emit a \texttt{NXT\_STEP}.  Unlike other common methods where each symbol will represent all the notes inside a time step, we decompose them into multiple symbols and the advancement of the time step is explicitly expressed using the symbol \texttt{NXT\_STEP}. 

\subsubsection{Note Feature}

With the previous encoding mechanism, we can encode any of the Beatles' songs into a sequence $S=\{{S_i}\}_{i=0}^{N}$. Here $S_i \in \mathcal{S}$ in which $\mathcal{S}$ is the set of all the possible symbols.  We have $|\mathcal{S}| = |T|*2 + 2$, where $T$ is the set of possible pitches.

Because the training data is limited, it is helpful to incorporate additional features for each symbol to help the neural network learn the theory and patterns of the music.  We pair each symbol $S_i$ with its feature $F_i$ when we feed the encoded sequence into the RNN.  We designed two features for BandNet, i.e., $F_i=(B_i, G_i)$.  The feature $B_i \in \{0,1\}^5 $ contains the beat information.  $B_i=1$ if and only if the global time step of $i$th symbol is a multiple of $2^i$.  We find that this feature is helpful for the RNN to keep the style of the chord channel consistent inside a measure.  The second feature $G_i \in \{0,1\}$ represents whether the melody will be generated at the current time step.  Without this feature, we find that sometimes BandNet will not generate a vocal melody due to silences in the melody channel of the training data (usually because of an instrumental or guitar solo section).  By setting this variable to one or zero, we can easily control whether we want to generate the vocal part in a given section of music.

\subsection{Network Structure}
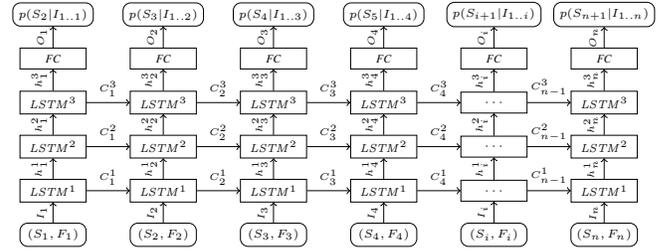
\begin{figure}
    \centering
    \resizebox{\linewidth}{!}{%
    \begin{tikzpicture}[yscale=0.8,xscale=2.0,font=\scriptsize,every node/.style={draw,rectangle, minimum size=0mm, minimum height=1.2em, minimum width=1.2cm}]
    \foreach \x in {1,...,6} {%
        \foreach \y in {0,1,2,3,4,5} {
            \ifnum\x=5
                \ifnum\y=0
                    \node[rounded corners] (L\x\y) at (\x, \y) {$(S_i, F_i)$};
                \else
                    \ifnum\y=5
                        \node[rounded corners] (L\x\y) at (\x, \y) {$p(S_{i+1}|I_{1..i})$};
                    \else
                        \ifnum\y=4
                            \node (L\x\y) at (\x, \y) {$\textit{FC}$};
                        \else
                            \node (L\x\y) at (\x, \y) {$\cdots$};
                        \fi
                    \fi
                \fi
            \else
                \ifnum\x=6
                    \def\xt{n}
                    \def\xx{n+1}
                \else
                    \pgfmathtruncatemacro{\xt}{\x}
                    \pgfmathtruncatemacro{\xx}{\x+1}
                \fi

                \ifnum\y=0
                    \node[rounded corners] (L\x\y) at (\x, \y) {$(S_{\xt}, F_{\xt})$};
                \else\ifnum\y=4
                    \node (L\x\y) at (\x, \y) {$\textit{FC}$};
                \else\ifnum\y=5
                    \node[rounded corners] (L\x\y) at (\x, \y) {$p(S_{\xx} | I_{1..{\xt}})$};
                \else
                    \node (L\x\y) at (\x, \y) {$\mathit{LSTM}^\y$};
                \fi
                \fi
                \fi
            \fi
        }
    }

    \foreach \x in {1,2,3,4,5,6} {%
        \ifnum\x=6
            \def\xt{n}
            \def\xx{n+1}
        \else
            \ifnum\x=5
                \def\xt{i}
                \def\xx{i+1}
            \else
                \pgfmathtruncatemacro{\xt}{\x}
                \pgfmathtruncatemacro{\xx}{\x+1}
            \fi
        \fi
        \path[every node/.style={sloped,anchor=south},font=\tiny] (L\x0) edge[->] node {$I_{\xt}$} (L\x1);
        \path[every node/.style={sloped,anchor=south},font=\tiny] (L\x1) edge[->] node {$h_{\xt}^1$} (L\x2);
        \path[every node/.style={sloped,anchor=south},font=\tiny] (L\x2) edge[->] node {$h_{\xt}^2$} (L\x3);
        \path[every node/.style={sloped,anchor=south},font=\tiny] (L\x3) edge[->] node {$h_{\xt}^3$} (L\x4);
        \path[every node/.style={sloped,anchor=south},font=\tiny] (L\x4) edge[->] node {$O_{\xt}$} (L\x5);
    }

    \foreach \x in {1,2,3,4,5} {%
        \pgfmathtruncatemacro{\xx}{\x+1}
        \ifnum\x=5
            \def\xt{n-1}
        \else
            \pgfmathtruncatemacro{\xt}{\x}
            \pgfmathtruncatemacro{\xx}{\x+1}
        \fi
        \path[every node/.style={sloped,anchor=south},font=\tiny] (L\x1) edge[->] node {$C_{\xt}^1$} (L\xx1);
        \path[every node/.style={sloped,anchor=south},font=\tiny] (L\x2) edge[->] node {$C_{\xt}^2$} (L\xx2);
        \path[every node/.style={sloped,anchor=south},font=\tiny] (L\x3) edge[->] node {$C_{\xt}^3$} (L\xx3);
    }
\end{tikzpicture}
}
    \caption{\small
    A diagram showing how an unrolled 3-layer LSTM-RNN works for music composition.  Here, symbol $S_i$ and feature $F_i$ are encoded to the vector $I_i$.  $\mathit{LSTM}^j$ represents an LSTM cell in the $j$th layer.  Cells in the same layer share the same parameter.  $C_i^j$ and $h_i^j$ are the cell state and hidden state of the $i$th cell in the $j$th layer.  $\mathit{FC}$ represents a fully-connected layer and its output $O_i$ is fed into a softmax function to produce a distribution over all the possible symbols.
    }
    \label{fig:rnn}
\end{figure}

Figure \ref{fig:rnn} shows how a classical multi-layer LSTM-RNN \cite{hochreiter1997long} models the probabilistic distribution of the symbol sequence.  At the bottom layer, each LSTM cell takes the symbol $S_i$ in its one-hot vector form together with the corresponding binary feature vector $F_i$ as its input $I_i$.  These LSTM cells are chained so that they will apply nonlinear transformations to the previous cell state $C_{i-1}^1$ and input $I_i$ and produce the current hidden state $h_i^1$ and cell state $C_{i}^1$.  In order to increase the nonlinearity of the model, we make the network deep by stacking multiple layers of LSTM cells.  Starting from the second layer, each cell will take the hidden state from the previous layer as input.  Finally, we apply a linear transformation to the hidden states in the last layer with softmax to compute the conditional probability $P_{\Theta}(S_{i+1} \,|\, I_{\{1\cdots i\}})$, where $\Theta$ contains the parameters of the network.  We use BPTT \cite{mozer1989focused} to find the parameters that locally maximizes the likelihood of the training data.

\subsection{Keeping Notes in the Key}

The melody channel generated by our model occasionally contained unexpected notes.  We found that many of these notes are dissonant because they are not in the key of the music.  We speculate that this is because the Beatles often used  notes in their music that deviated from conventional practices of other popular music. These notes may work well under some conditions, but the amount of data does not allow our neural network to learn how to use these notes in the right context.  Therefore, in order to improve the quality of our music, it is reasonable to filter them out in BandNet, i.e., restricting the notes that are not in the song's key during the generating stage. This can be achieved by applying a mask to the probability distributions returned by the neural network and re-normalizing them so that they all sum to unity.
%
%

\subsection{Generating a Complete Song} \label{sec:song}

\begin{figure}
    \centering
    \includegraphics[width=\linewidth]{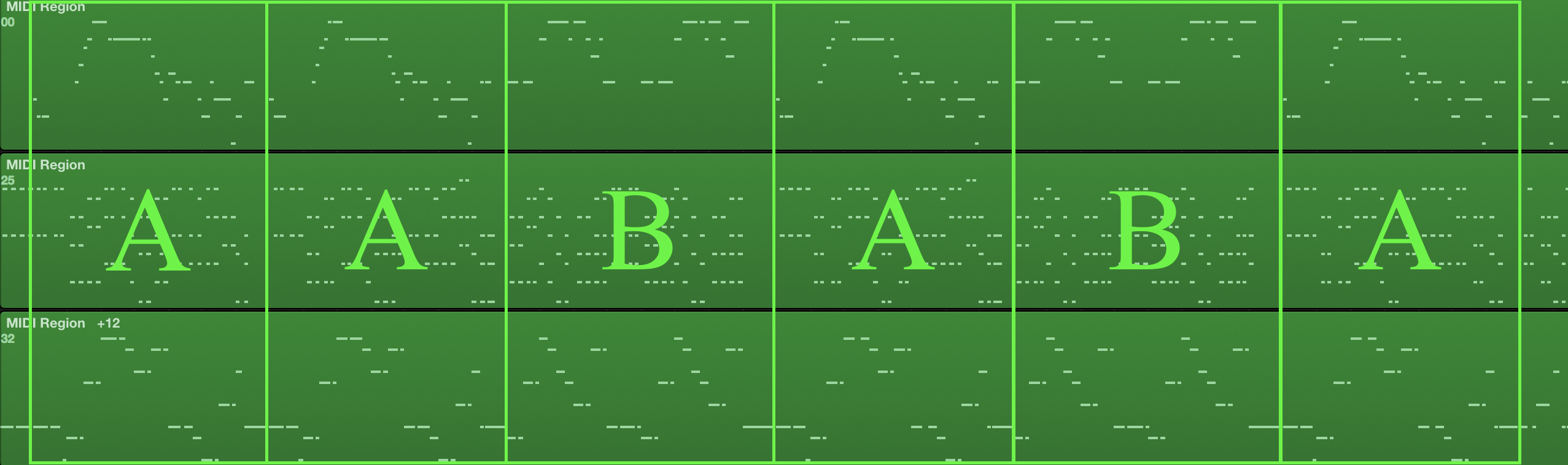}
    \caption{\small
    The piano roll of the song \emph{Yesterday (1965)}.  It has a song structure AABABA, whose sections are labeled in green in the Figure.  The channels from top to bottom are melody, chords, and bass line.
    }
    \label{fig:repeating}
\end{figure}

Most of the Beatles' music has a repetitive and sectional song structure.  Figure \ref{fig:repeating} shows an example of the structure in the song \emph{Yesterday (1965)}.  This song uses an \emph{AABABA} structure, where the A section is called the \textit{verse} and the B section is called the \textit{chorus}.  The verse section is repeated four times, with each repetition being exactly the same or having only minor differences.  It is hard for the RNN to learn this phenomenon because the distance between two sections is as long as eight measures, i.e., 128 time steps.  RNN normally cannot carry hundreds of symbols in its memory across a span of that long.  Folk-RNN \cite{sturm2015folk} used a data format called ABC notation that has an annotation for repeating sections so that they do not need to deal with this problem. We do not have such fine-level annotation in our dataset.  Instead, we use a template-based method to generate structured music.  Users of BandNet will first select a predefined song structure template, e.g., AABA or ABABCBB, and then BandNet can generate a clip for each section whose length can vary from 4 to 16 measures.  After that, we assemble the generated clips to form a complete song.  Because we do not model the drum pattern in this work, we assign a precomposed drum pattern for each section of music, which is beneficial as we can select different styles of drum patterns for different sections of the song.

The well-known DeepBach \cite{hadjeres2016deepbach} and BachBot \cite{liang2016bachbot} can generate a new harmony or re-harmonize an existing melody from a single instrument, i.e. piano. BandNet can generate a song with multiple instruments, e.g. guitar, keyboard, bass, and drum.  Because we do not have a melody to condition on, BandNet needs a short sequence of notes, also known as a \emph{seed}, to begin a section.  Although in theory it is possible not to condition on any seeds, we found that the resulting music was often unsatisfactory.  In order to avoid depending on a professional musician to compose note sequences as seeds, we adopt the following strategy:  First, we let BandNet generate long sequences of music without conditioning on any seeds.  Second, we can listen to these randomly generated segments and mark the clips that sound most compelling to us.  Third, we use these clips as seeds for BandNet to generate all the sections of the song.

\section{Experiments} \label{sec:exp}

\subsection{Settings}

We collected 183 Beatles MIDI songs from the Internet as our training dataset.  We removed 60 songs from the dataset because they were either divergent in musical style when compared with other Beatles' songs, or were missing important components such as a clear vocal melody or bass line.  We found that MIDI files in the wild can be messy.  For example, the chords may be divided across three channels in some MIDI files, while there can be up to eight channels used for instrumental decoration in others, which is not necessary for our purposes.  We cleaned this dataset by deleting the unnecessary channels and merging the fragmented channels.

Due to the number of songs that the Beatles composed, the size of our dataset is smaller compared to those used in the literature \cite{liang2016bachbot,chu2016song,yang2017midinet}, but we found that it is sufficient to train a reasonably good model.  Aside from its influence in popular music history, there are two reasons why we choose to use the Beatles' catalog as our training dataset: First, the style of the Beatles' music is relatively consistent when compared to other categories of pop music, and therefore it is easier for the RNN to learn its underlying structures.  Second, most of the Beatles' music contains the elements required by our music generation pipeline, such as distinct melody, chord, and bass parts, as well as repeating song structures, which can be missing in genres such as classical and folk music.  

The two most important parameters of the recurrent neural network were the dimension of LSTM cells and the number of layers.  We found that a 3-layer RNN in which each LSTM cell had 256 hidden units worked well in practice.  

Our implementation was based on Magenta \cite{magenta14} and Tensorflow \cite{tensorflow2015-whitepaper} for processing the MIDI files and training the RNN.  Because the number of parameters in our network was large, we applied dropout \cite{srivastava2014dropout} to alleviate overfitting.  We trained our model using the Adam optimizer \cite{kingma2014adam}, which is a variant of stochastic gradient descent that is not sensitive to the global learning rate.  We used 10\% songs in our dataset for cross validation and we stopped the training process when the error on the validation dataset no longer decreased.  During the training, we clipped the gradients so that their L2-norms were less than or equal to 1.  This technique was proposed in \cite{srivastava2014dropout} to prevent the gradient explosion problem.


\subsection{Quality Scoring by a Professional Composer}

 

In this section, a professional music composer evaluated the music generated by each subsequent version of BandNet. The composer gave two scores for each individual channel (melody, chords, and bass) based on their musical content and structure. The Content Quality (CQ) was defined as how well the notes and rhythms in the generated music function according to music theory principles consistent with the music of the Beatles, and the Structure Quality (SQ) was defined as to what extent the music sample exhibits an organizational structure.  All scores were given on a scale of 1 to 5.  In addition, we designed two overall scores to evaluate the overall quality of each multiple-channel song. The Averaged Content and Structure Quality (ACSQ) were calculated through averaging the CQs and SQs of all the channels, and the Group Synergy Quality (GSQ) score evaluated how well the individual channels work together to make a unified whole.


\begin{table}
\small
    \centering
    \resizebox{\linewidth}{!}{
        \begin{tabular}{|l|c|c|c|c|c|c||c|c|}
        \hline 
          &  \multicolumn{2}{|c|}{Melody} & \multicolumn{2}{|c|}{Chords} & \multicolumn{2}{|c||}{Bass} &  &   \\
        \hline
         & CQ & SQ & CQ & SQ & CQ & SQ & ACSQ & GSQ \\
         \hline
        MGT-M         & 2.60  & 2.70 & - & - & - & - & - & 2.65 \\
        MGT-P         & - & - & 3.20  & 2.50 & - & - & - & 2.85 \\
        BN            & 2.90  & 1.50 & 2.70 & 2.40 & 3.30 & 2.40 & 2.53 & 2.60 \\
        BN-S         & 2.90 & 2.50 & 3.05 & 2.90 & 3.20 & 3.20 & 2.96 & 2.95 \\
        BN-SB     & 2.90 & 3.40 & 2.85 & 3.25 & 3.30 & 3.25 & 3.16 & 3.10\\
        BN-SBK & 3.85 & 3.75 & 3.45 & 3.45 & 3.75 & 3.65 & 3.65  & 3.90 \\
        BEATLES & 4.45 & 4.80 & 4.20 & 4.75 & 4.40 & 4.95 & 4.59  & 4.65  \\
         \hline
    \end{tabular}
    }
    \caption{\small
    Results of a professional composer evaluating the quality of music generated by different models. \textbf{MGT-M}: Magenta's MelodyRNN, \textbf{MGT-P}: Magenta's PolyphonyRNN, \textbf{BN}: BandNet without note features, \textbf{BN-S}: BN with \textbf{s}ilence feature, \textbf{BN-SB}: BN-S with \textbf{b}eat feature, \textbf{BN-SBK}: BN-SB while keeping notes in the \textbf{k}ey, \textbf{BEATLES}: original Beatles' songs. The definitions of CQ, SQ, ACSQ, and GCQ can be found in Section 3.2.
    }
    \label{tab:t1}
\end{table}

The results are shown in Table \ref{tab:t1}. The score was an average across five songs under each setting.  We found that model BN was on par with Magenta's melody and polyphony generators \cite{magenta14} in terms of content and structure scores, which is reasonable because models from Magenta were designed to model melody and chords (as in polyphonic music) separately, and modeling them jointly in the case of BandNet would not improve the score of each individual channel.  After introducing the silence feature, the GSQ of BandNet increased from 2.6 to 2.95 because we were able to exclude unusual silences in the melody. By adding the beat feature, BandNet continued to receive rewards in SQs for the melody and chord channels; a possible explanation for this is that the beat feature gave the RNN measure and section information, which helped it learn the structure of the music more efficiently.  Both of these features also improved GSQs, as the normalization of each individual channel also improved the alignment between individual parts.  Finally, the greatest improvement in both metrics was from the key restriction feature. This significantly improved the CQs of individual channels by removing ``wrong'' notes, and also improved SQs and GSQs by reducing the amount of notes that were dissonant with one another across individual channels.


\subsection{Subjective Listening}

We also conducted a subjective listening experiment to evaluate the quality of our generated songs from the perspective of amateurs.  We received 17 responses in this user study: 16 said that they had never received formal musical training.  In this test, we asked users to listen to 15 songs. All of the songs were in AABA structure and each section had a length of 8 measures.  The first 5 songs, labeled as group A, were composed by BandNet using randomly generated seeds; the next 5 songs, labeled as group B, were composed by BandNet using professionally composed seeds.  Each seed was 2 measures in length, with BandNet generating the remaining 6-measure clip for each section. Songs in group A and B were generated randomly without human selection.  The last 5 songs, labeled as group C, were relatively unknown Beatles' songs, with the intention that listeners had likely never heard them before. We shuffled the order of the songs so that listeners could not guess whether a song was composed by BandNet prior to listening. We also modified the drum patterns for the group C Beatles' songs, so that listeners could not distinguish them from BandNet-composed songs based on differences in the drum pattern.

At the beginning, we asked subjects to listen to 5 well-known songs by the Beatles, such as \emph{I Want to Hold Your Hand (1964)}, in order to familiarize them with the Beatles’ musical style.   Next, we asked them to listen to the 15 songs mentioned above and to answer the following 4 questions for each song:

\begin{compactitem}
  \item[{Q1}:] Have you heard this song before?
  \item[{Q2}:] Does it sound similar to the music of the Beatles?
  \item[{Q3}:] How likely is it that this music was professionally composed?
  \item[{Q4}:] How interesting is this music?
\end{compactitem}

\begin{figure}
    \centering
    \begin{tikzpicture}
    \begin{axis}[
        boxplot/draw direction=y,
        ylabel={Score (higher is better)},
        boxplot={
            draw position={1/4 + 1*floor(\plotnumofactualtype/3) + 1/4*mod(\plotnumofactualtype,3)},
            whisker range=10000,
            box extend=0.2,
        },
        grid,
        x=22mm,
        height=4cm,
        ymin=1.5,
        ymax=4.5,
        xtick={0,1,2,...,10},
        ytick={1,2,...,5},
        x tick label as interval,
        xticklabels={%
            {Style Similarity},%
            {Professional Sounding},%
            {Interestingness},%
        },
        label style={font=\scriptsize},
        tick label style={font=\scriptsize},
        legend style={font={\fontsize{4}{4}\selectfont},row sep=-3pt},
    cycle list={{red},{blue},{brown}},
    name=border,
    legend entries = {{\,Group A: BandNet, Generated Seeds}, \,{Group B: BandNet, Professional Seeds}, \,Group C: The Beatles' Music},
    legend cell align=left,
    legend to name={legend},
]

\addplot
table[row sep=\\,y index=0] {
data\\
2.75\\
2.8125\\
3.21875\\
3.21875\\
3.375\\
};

\addplot
table[row sep=\\,y index=0] {
data\\
2.96875\\
3.1875\\
3.0000\\
3\\
2.9375\\
};

\addplot
table[row sep=\\,y index=0] {
data\\
3.433333\\
3.076923\\
3.375\\
3.166667\\
3.05\\
};

\addplot
table[row sep=\\,y index=0] {
data\\
3.1875\\
3.53125\\
3.53125\\
3.34375\\
2.84375\\
};

\addplot
table[row sep=\\,y index=0] {
data\\
3.53125\\
3.28125\\
3.13333\\
3.2875\\
3.29875\\
};

\addplot
table[row sep=\\,y index=0] {
data\\
3.7333333\\
3.730769\\
3.65625\\
3.82142\\
3.454545\\
};

\addplot
table[row sep=\\,y index=0] {
data\\
3.09375\\
3.21875\\
3.625\\
3.03125\\
2.96666\\
};

\addplot
table[row sep=\\,y index=0] {
data\\
3.53125\\
3.125\\
3.2\\
2.6875\\
3.1\\
};
\addplot
table[row sep=\\,y index=0] {
data\\
3.7333333\\
3.615384\\
3.59375\\
3.933333\\
3.5\\
};
\end{axis}

\node[above left] at (border.south east) {\ref{legend}};

\end{tikzpicture}
    \caption{\small
    Result of a user study that evaluates the performance of different ways to generate music.  The $x$-axis represents the sources of the music and the $y$-axis represents the score.  The box plot shows the distribution of the average score of each song rated by the listener.
    }
    \label{fig:t2}
\end{figure}
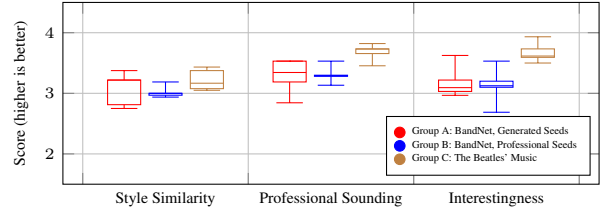

We asked listeners to only choose between ``Yes, definitely!'' and ``No/Not sure'' in Q1; if they answered ``Yes'', we  removed their scoring of that song from our results.  This is because a subject may be biased to give a song a higher score if he had heard it song before.  For Q2, Q3, and Q4, we let users grade each song using a scale from 1 to 5 with an increment of 0.5.  Figure \ref{fig:t2} shows the distribution of those scores from 17 responses.  The labels in the horizontal axis, Style Similarity, Professional Sounding, and Interestingness correspond to Q2, Q3, and Q4, respectively.  Each sample in the box plot represents the average score over 17 responses to a question for a particular song.

For Q1,  about 13.3\% of responses indicated that they had heard the authentic Beatles' songs before, while the percentages were only 0\% and 1.3\% for BandNet-generated songs using automatically-generated seeds and professional seeds, respectively.  This could be an indicator showing that we did not overfit the training data and just replicated some clips from the original Beatles' music.  For the rest of the questions, we found that the authentic Beatles' songs constantly outperformed the BandNet-generated songs, but only by a small margin.  In particular, the average Style Similarity scores for songs in group A, B, and C are 3.08, 3.02, and 3.22, respectively.  The score difference of Q2 between the authentic and generated songs was less than 0.202, which showed that BandNet was able to imitate the style of the Beatles relatively well.  The average Professional Sounding scores were 3.29, 3.16, and 3.68, and the average Interestingness scores were 3.19, 3.13, and 3.68 for songs in group A, B, and C, respectively. The score gaps of Q3 and Q4 between authentic and generated songs were approximately 0.5. 
The musical knowledge that BandNet learned came primarily from The Beatles, and in theory may be difficult for a RNN-based machine learning algorithm to generate more professional and interesting music than The Beatles. Concerning the seeds used in generation, our experiments have shown that using professionally-composed seeds did not have a significant advantage over selecting from randomly-generated seeds in terms of subjective listening evaluation. This means that we may no longer need a composer in the loop for generating a complete song and an amateur would be able to ``compose'' a Beatles-style song without the guide of a professional by using BandNet. 

%

\section{Conclusions}

We have proposed a RNN-based, multi-instrument MIDI music composition machine, which can learn musical knowledge from existing Beatles' music and automatically generate music in the style of the Beatles with little human intervention. We also integrated expert knowledge into the data-driven based learning process. Our approach has  proved to be effective by both professional evaluation and subjective listening tests.

\nocite{*}


\bibliographystyle{plain}
\bibliography{references}

\begin{thebibliography}{10}

\bibitem{tensorflow2015-whitepaper}
Mart\'{\i}n Abadi, Ashish Agarwal, Paul Barham, Eugene Brevdo, Zhifeng Chen,
  Craig Citro, Greg~S. Corrado, Andy Davis, Jeffrey Dean, Matthieu Devin,
  Sanjay Ghemawat, Ian Goodfellow, Andrew Harp, Geoffrey Irving, Michael Isard,
  Yangqing Jia, Rafal Jozefowicz, Lukasz Kaiser, Manjunath Kudlur, Josh
  Levenberg, Dan Man\'{e}, Rajat Monga, Sherry Moore, Derek Murray, Chris Olah,
  Mike Schuster, Jonathon Shlens, Benoit Steiner, Ilya Sutskever, Kunal Talwar,
  Paul Tucker, Vincent Vanhoucke, Vijay Vasudevan, Fernanda Vi\'{e}gas, Oriol
  Vinyals, Pete Warden, Martin Wattenberg, Martin Wicke, Yuan Yu, and Xiaoqiang
  Zheng.
\newblock {TensorFlow}: Large-scale machine learning on heterogeneous systems,
  2015.
\newblock Software available from tensorflow.org.

\bibitem{allan2005harmonising}
Moray Allan and Christopher Williams.
\newblock Harmonising chorales by probabilistic inference.
\newblock In {\em Advances in neural information processing systems}, pages
  25--32, 2005.

\bibitem{boulanger2012Modeling}
Nicolas Boulanger{-}Lewandowski, Yoshua Bengio, and Pascal Vincent.
\newblock Modeling temporal dependencies in high-dimensional sequences:
  Application to polyphonic music generation and transcription.
\newblock In {\em Proceedings of the 29th International Conference on Machine
  Learning, {ICML} 2012, Edinburgh, Scotland, UK, June 26 - July 1, 2012},
  2012.

\bibitem{magenta14}
Google Brain.
\newblock Magenta.
\newblock \url{https://magenta.tensorflow.org/}, 2000--2004.

\bibitem{chu2016song}
Hang Chu, Raquel Urtasun, and Sanja Fidler.
\newblock Song from {PI}: A musically plausible network for pop music
  generation.
\newblock {\em arXiv preprint arXiv:1611.03477}, 2016.

\bibitem{cuthbert2010music21}
Michael~Scott Cuthbert and Christopher Ariza.
\newblock music21: A toolkit for computer-aided musicology and symbolic music
  data.
\newblock 2010.

\bibitem{ebciouglu1988expert}
Kemal Ebcio{\u{g}}lu.
\newblock An expert system for harmonizing four-part chorales.
\newblock {\em Computer Music Journal}, 12(3):43--51, 1988.

\bibitem{eppe2015computational}
Manfred Eppe, Roberto Confalonieri, Ewen Maclean, Maximos Kaliakatsos, Emilios
  Cambouropoulos, Marco Schorlemmer, Mihai Codescu, and K~K{\"u}hnberger.
\newblock Computational invention of cadences and chord progressions by
  conceptual chord-blending.
\newblock IJCAI'15 Proceedings of the 24th International Conference on
  Artificial Intelligence, 2015.

\bibitem{hadjeres2016deepbach}
Ga{\"e}tan Hadjeres and Fran{\c{c}}ois Pachet.
\newblock {DeepBach}: a steerable model for {Bach} chorales generation.
\newblock In {\em Proceedings of the 34th International Conference on Machine
  Learning}, 2017.

\bibitem{hadjeres2016style}
Ga{\"e}tan Hadjeres, Jason Sakellariou, and Fran{\c{c}}ois Pachet.
\newblock Style imitation and chord invention in polyphonic music with
  exponential families.
\newblock {\em arXiv preprint arXiv:1609.05152}, 2016.

\bibitem{hild1992harmonet}
Hermann Hild, Johannes Feulner, and Wolfram Menzel.
\newblock Harmonet: A neural net for harmonizing chorales in the style of js
  bach.
\newblock In {\em Advances in neural information processing systems}, pages
  267--274, 1992.

\bibitem{hiller1979experimental}
Lejaren~Arthur Hiller and Leonard~M Isaacson.
\newblock {\em Experimental Music; Composition with an electronic computer}.
\newblock Greenwood Publishing Group Inc., 1979.

\bibitem{hochreiter1997long}
Sepp Hochreiter and J{\"u}rgen Schmidhuber.
\newblock Long short-term memory.
\newblock {\em Neural computation}, 9(8):1735--1780, 1997.

\bibitem{huang2016counterpoint}
Cheng-Zhi~Anna Huang, Tim Cooijmans, Adam Roberts, Aaron Courville, and Douglas
  Eck.
\newblock Counterpoint by convolution.
\newblock 2016.

\bibitem{kaliakatsos2014probabilistic}
Maximos Kaliakatsos-Papakostas and Emilios Cambouropoulos.
\newblock Probabilistic harmonization with fixed intermediate chord
  constraints.
\newblock In {\em ICMC}, 2014.

\bibitem{kingma2014adam}
Diederik Kingma and Jimmy Ba.
\newblock Adam: A method for stochastic optimization.
\newblock {\em arXiv preprint arXiv:1412.6980}, 2014.

\bibitem{liang2016bachbot}
Feynman Liang, Mark Gotham, Matthew Johnson, and Jamie Shotton.
\newblock {BachBot}: Automatic composition in the style of bach chorales.
\newblock In {\em Proceedings of the 18th International Society for Music
  Information Retrieval Conference (ISMIR’2017)}, 2017.

\bibitem{makris2017combining}
Dimos Makris, Maximos Kaliakatsos-Papakostas, Ioannis Karydis, and Katia~Lida
  Kermanidis.
\newblock Combining {LSTM} and feed forward neural networks for conditional
  rhythm composition.
\newblock In {\em International Conference on Engineering Applications of
  Neural Networks}, pages 570--582. Springer, 2017.

\bibitem{mozer1989focused}
Michael~C Mozer.
\newblock A focused back-propagation algorithm for temporal pattern
  recognition.
\newblock {\em Complex systems}, 3(4):349--381, 1989.

\bibitem{papadopoulos2016assisted}
Alexandre Papadopoulos, Pierre Roy, and Fran{\c{c}}ois Pachet.
\newblock Assisted lead sheet composition using {FlowComposer}.
\newblock In {\em International Conference on Principles and Practice of
  Constraint Programming}, pages 769--785. Springer, 2016.

\bibitem{pascanu2013difficulty}
Razvan Pascanu, Tomas Mikolov, and Yoshua Bengio.
\newblock On the difficulty of training recurrent neural networks.
\newblock In {\em International Conference on Machine Learning}, pages
  1310--1318, 2013.

\bibitem{quick2014kulitta}
Donya Quick.
\newblock {\em Kulitta: A framework for automated music composition}.
\newblock Yale University, 2014.

\bibitem{srivastava2014dropout}
Nitish Srivastava, Geoffrey~E Hinton, Alex Krizhevsky, Ilya Sutskever, and
  Ruslan Salakhutdinov.
\newblock Dropout: a simple way to prevent neural networks from overfitting.
\newblock {\em Journal of machine learning research}, 15(1):1929--1958, 2014.

\bibitem{sturm2015folk}
Bob Sturm, Joao~Felipe Santos, and Iryna Korshunova.
\newblock Folk music style modelling by recurrent neural networks with long
  short term memory units.
\newblock In {\em 16th International Society for Music Information Retrieval
  Conference}, 2015.

\bibitem{whorley2013multiple}
Raymond~P Whorley, Geraint~A Wiggins, Christophe Rhodes, and Marcus~T Pearce.
\newblock Multiple viewpoint systems: Time complexity and the construction of
  domains for complex musical viewpoints in the harmonization problem.
\newblock {\em Journal of New Music Research}, 42(3):237--266, 2013.

\bibitem{yang2017midinet}
Li-Chia Yang, Szu-Yu Chou, and Yi-Hsuan Yang.
\newblock {MidiNet}: A convolutional generative adversarial network for
  symbolic-domain music generation.
\newblock In {\em Proceedings of the 18th International Society for Music
  Information Retrieval Conference (ISMIR’2017), Suzhou, China}, 2017.

\end{thebibliography}

\end{document}